# TME-BNA: Temporal Motif-Preserving Network Embedding with Bicomponent Neighbor Aggregation

LING CHEN*

Zhejiang University

DA WANG

Zhejiang University

DANDAN LYU

Zhejiang University

XING TANG

Zhejiang University

HONGYU SHI

Zhejiang University

Evolving temporal networks serve as the abstractions of many real-life dynamic systems, e.g., social network and e-commerce. The purpose of temporal network embedding is to map each node to a time-evolving low-dimension vector for downstream tasks, e.g., link prediction and node classification. The difficulty of temporal network embedding lies in how to utilize the topology and time information jointly to capture the evolution of a temporal network. In response to this challenge, we propose a temporal motif-preserving network embedding method with bicomponent neighbor aggregation, named TME-BNA. Considering that temporal motifs are essential to the understanding of topology laws and functional properties of a temporal network, TME-BNA constructs additional edge features based on temporal motifs to explicitly utilize complex topology with time information. In order to capture the topology dynamics of nodes, TME-BNA utilizes Graph Neural Networks (GNNs) to aggregate the historical and current neighbors respectively according to the timestamps of connected edges. Experiments are conducted on three public temporal network datasets, and the results show the effectiveness of TME-BNA.

* This work was funded by the National Key Research and Development Program of China (No. 2018YFB0505000).
Authors' addresses: L. Chen (Corresponding author), College of Computer Science and Technology, Alibaba-Zhejiang University Joint Research Institute of Frontier Technologies, Zhejiang University, Hangzhou 310027, China; email: lingchen@cs.zju.edu.cn; D. Wang, D. Lyu, X. Tang, H. Shi, College of Computer Science and Technology, Zhejiang University, Hangzhou 310027, China; emails: {wangda9655, revaludo, tangxing, shihongyu}@cs.zju.edu.cn.

# 1 INTRODUCTION

Many systems in real life, e.g., social network [1] and e-commerce, can be modeled as temporal networks that change over time [2]. In a temporal network, the interacting objects are regarded as nodes and the time-stamped interactions are regarded as edges between nodes. As time goes by, new nodes and edges will appear in the temporal network constantly. Figure 1 shows an example of the temporal network in a social network. Users are regarded as nodes in the temporal network and the behaviors of sending messages between users, i.e., interactions, are regarded as edges. Temporal network embedding aims at learning the low-dimension representations of nodes changing over time, which are very useful for a variety of downstream tasks [3-6], e.g., friend recommendation in social network and item recommendation in e-commerce.

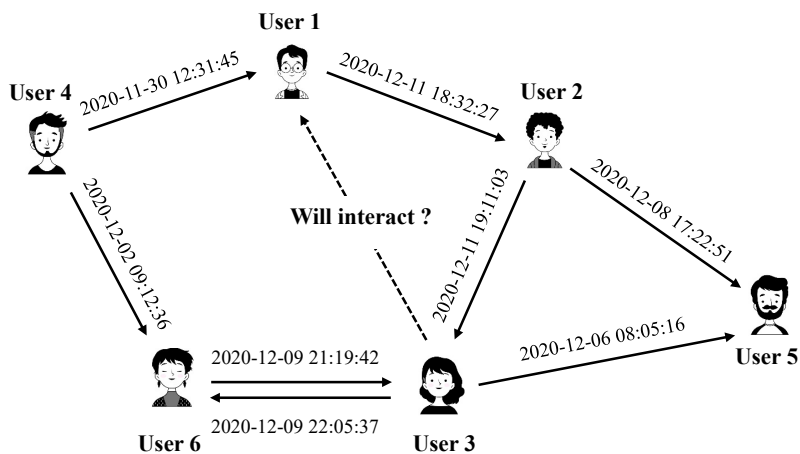

Figure 1: An example of a temporal network.

Existing temporal network embedding methods can be mainly divided into two categories: discrete-time dynamic graph embedding methods and continuous-time dynamic graph embedding methods [7]. Discrete-time dynamic graph embedding methods split the whole temporal network into a sequence of network snapshots by a fixed time window. Early studies impose smoothness constraints on the nodes over adjacent network snapshots to model the evolution of the temporal network [8, 9], which cannot capture the abrupt changes. With the development of deep neural networks in modeling sequence data, some methods exploit neural networks to capture the sequential dependencies of network snapshots [6, 10, 11]. These methods are sensitive to the choice of the time window, and it is difficult for them to capture the fine-grained dynamics of a temporal network.

Recently, continuous-time dynamic graph embedding methods that deal with interactions directly become promising temporal network embedding methods, which can preserve fine-grained time information. Early studies process the interacting nodes in chronological order but ignore the valuable topology of nodes [4, 12, 13]. In order to capture the topology dynamics of a temporal network, some methods learn node representations considering the neighbor influence of nodes. However, these methods only use one-hop neighbors [14, 15] or aggregate multi-hop neighbors layer by layer [16-18], which are unable to explicitly utilize the complex topology of a temporal network. In addition, existing multi-hop based methods aggregate neighbors in the descending chronological order, which cannot capture the recent changes of multi-hop neighbors. As shown in Figure 1, user 1 sent a message to user 2 followed by the immediate interaction between user 2 and user 3. In this situation, it is very likely that user 3 will interact with user 1 later. However, when aggregating the 2-hop neighbors of user 1, existing methods abandon user 3, as the interaction between user 2 and user 3 is later than user 1 and user 2, which neglects the interaction effect we mentioned above.

To address the aforementioned problems, we propose a temporal motif-preserving network embedding method with bicomponent neighbor aggregation (TME-BNA). TME-BNA models the complex topology and time information in a temporal network jointly, and leverages neighbor information in different time periods. The main contributions of this paper are as follows:

1) We propose an additional edge feature construction method based on temporal motifs. Since temporal motifs are induced subgraphs on the sequences of edges that reflect the evolution patterns of a temporal network [19], introducing the corresponding edge feature can explicitly integrate complex topology with time information.

2) We introduce a bicomponent neighbor aggregation method based on multi-head attention. TME-BNA aggregates the historical and current neighbors of nodes respectively according to the timestamps of connected edges through the multi-head attention mechanism, which can capture the recent changes of multi-hop neighbors in a temporal network.

3) We evaluate TME-BNA on three public temporal network datasets and make comparisons with the state-of-the-art temporal network embedding methods. Experimental results show that TME-BNA can achieve superior performance.

The rest of this paper is organized as follows. Section 2 reviews the related work. Section 3 gives the preliminaries of this paper and defines the problem formally. Section 4 introduces the proposed method TME-BNA in detail. Section 5 presents the experimental settings and corresponding results. Finally, Section 6 concludes the paper and gives a brief discussion of the future work.

## 2 RELATED WORK

In this section, we review the previous works related to this paper, including discrete-time dynamic graph embedding, continuous-time dynamic graph embedding, and motif-preserving network embedding.

### 2.1 Discrete-Time Dynamic Graph Embedding

Discrete-time dynamic graph embedding aims to learn node representations within consecutive snapshots that are split by a fixed time window. In order to model the dynamics of nodes in a temporal network, early studies usually impose smoothness constraints on the nodes. Zhu et al. [8] proposed a fast BCGD method, which performs matrix decomposition with smoothing constraints on the decomposition results of adjacent snapshots. Dunlavy et al. [9] proposed to perform tensor factorizations on the network snapshot sequence and utilize the Holt-Winter method to learn the influence of time factors. It is obvious that these methods are unable to capture the abrupt changes in a temporal network.

Recurrent Neural Networks (RNNs) (including its variants) and Transformers are deep neural networks that specialize in processing sequence data [20, 21]. Recently, more and more researchers adopt these methods to exploit sequential effects in a temporal network. Manessi et al. [10] proposed to utilize Graph Neural Networks (GNNs) [22] on each network snapshot to obtain the embedding sequences of nodes, and Long

Short-Term Memory Networks (LSTMs) [23] to model their dynamics. Sankar et al. [6] proposed DySAT, which replaces LSTMs with the self-attention mechanism [20] to exploit the long-term dependence in the network snapshot sequence. Chen et al. [24] proposed DACHA, which introduces a dual graph convolution network to capture the influence of both entities and historical relations, and utilizes self-attentive encoder to model temporal dependence in temporal knowledge graphs. Pareja et al. [11] proposed EvolveGCN, which introduces RNNs to model the parameter dynamics of the GNNs used in adjacent network snapshots. However, these methods need to choice a time window manually, which cannot model the fine-grained dynamics of a temporal network.

### 2.2 Continuous-Time Dynamic Graph Embedding

Recently, continuous-time dynamic graph embedding that can preserve fine-grained time information has attracted more and more attention. Some studies only consider the two interacting nodes involved in each interaction when processing the streaming interaction sequence, which ignore the valuable topology of nodes. Trivedi et al. [12] proposed Know-evolve, which introduces RNNs to update node representations according to the interaction sequence, then predicts the involving nodes and timestamp of the next interaction based on the temporal point process [25]. Kumar et al. [4] proposed JODIE, which adopts RNNs to update the representations of related nodes chronologically according to the interaction sequence, and predicts the node embedding of the next interaction directly. Similarly, Zhang et al. [13] proposed TigeCMN, which uses memory network [26] to update node representations sequentially based on the attention mechanism. The problem of these methods is that they only update the representations of related nodes according to the interaction sequence, which ignores the topology of nodes.

Due to the effectiveness of introducing the topology of nodes to capture their dynamics, recent methods take neighbor information into consideration. Zou et al. [14] proposed HTNE, which assigns different weights to neighbors based on the attention mechanism, then integrates the Hawkes process to model the influence of historical neighbors. Trivedi et al. [15] proposed DyRep, which updates the representations of interacting nodes and their one-hop neighbors with RNNs, then predicts the future interactions between nodes based on the temporal point process. Xu et al. [16] proposed TGAT, which maps interaction time to a high-dimension vector as an edge feature, then utilizes GNNs to aggregate multi-hop neighbors. Rossi et al. [17] proposed TGNs, which introduces a memory module to update node representations before the neighbor aggregating phase of TGAT. Wang et al. [18] proposed APAN, which updates node representations through a multi-head attention based synchronous module and propagates the influence to neighbors through an asynchronous module. Although the above methods take neighbors into consideration, there are two major shortcomings. First, they cannot utilize complex topology with time information explicitly. Second, the neighbor message passing directions follow the descending chronological order, which cannot capture the recent changes of multi-hop neighbors.

### 2.3 Motif-Preserving Network Embedding

A motif refers to a subgraph that appears more frequently in the real network than a random network, which is a common tool for analyzing complex networks [27]. Recently, more and more researchers introduce motifs into network embedding. Dareddy et al. [28] proposed Motif2vec, which transforms the original network based on motifs, then utilizes the skip-gram model [29] to learn node representations with random walks generated on the converted network. Lee et al. [30] proposed MCN, which constructs new adjacency matrices based on motifs, and uses the attention mechanism to model the influence of different motifs. Bouritsas et al. [31] proposed GSN, which introduces isomorphic graph counts as the additional features of nodes and edges to model complex topology in a static network. These studies show that integrating motifs is an effective approach to static network embedding, but there is no corresponding attempt in temporal network embedding.

## 3 PRELIMINARIES

In this section, we give the definitions of basic concepts and then formally define the problem.

**Definition 1: Temporal network**. A temporal network is a sequence of interactions sorted by interaction time, which can be denoted as $\mathcal{G} = \{e_1, e_2, \dots, e_N\}$. A triple $e = (v_i, v_j, t)$ indicates node $v_i \in V$ and node $v_j \in V$ has an interaction $e \in E$ at time $t$, where $V$ and $E$ are the sets of nodes and interactions in the temporal network, respectively. Figure 2(a) shows a temporal graph $\mathcal{G}_1$ with 4 nodes and 7 interactions.

**Definition 2: $\delta$-temporal motif**. A temporal motif represents a small subgraph pattern that appears frequently in a temporal network. In order to distinguish it from specific subgraph instances in the temporal network, an interaction in the temporal motif is denoted as $c = (u_i, u_j, t')$, which means $u_i$ and $u_j$ have an interaction $c$ at time $t'$. The $\delta$-temporal motif can be defined as $M = \{c_1, c_2, \dots, c_L\}$, where all the edges are sorted chronologically within $\delta$ duration [19]. Figure 2(b) shows an example of a 3-node, 3-egde $\delta$-temporal motif $M_1$.

**Definition 3: $\delta$-temporal motif instance**. For the $\delta$-temporal motif $M = \{c_1, c_2, \dots, c_L\}$, a sequence $S = \{e_1, e_2, \dots, e_L\}$ consisting of $L$ interactions in the temporal network is an instance of $M$ if it satisfies the following two conditions: (1) $M$ and $S$ can derive the same topology and the edges are connected in the same order; (2) All interactions in the sequence $S$ occur within $\delta$ duration. Figure 2(c) gives two instances of $M_1$ in the temporal network $\mathcal{G}_1$.

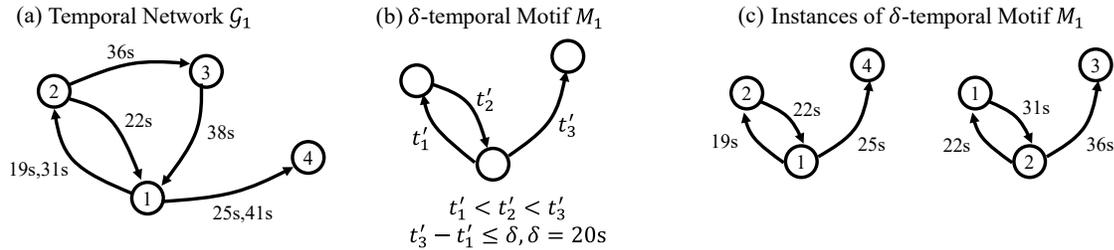

Figure 2: An example of a $\delta$-temporal motif and its instances.

**Definition 4: Temporal network embedding**. Given a temporal network $\mathcal{G}$, our goal is to learn a function $f: V \to \mathbb{R}^d$ to map each node in the temporal network to a low-dimension embedding, where $d$ is the number of embedding dimension.

## 4 METHODOLOGY

This section gives the detailed description of the proposed method TME-BNA. We first give the framework of TME-BNA, and then present the details of the modules in the proposed method.

## 4.1 Framework

The overall framework of TME-BNA is shown in Figure 3, which consists of four parts: 1) Node memory update; 2) Temporal motif based edge feature construction; 3) Bicomponent neighbor aggregation; 4) Link prediction.

The node memory update module utilizes RNNs to update the representations of the nodes involved in the interactions to model the dynamics of nodes. The temporal motif based edge feature construction module counts all the instances of the temporal motifs for each edge. The bicomponent neighbor aggregation module incorporates temporal motif based edge features to explicitly utilize the complex topology with time information, and integrates the multi-head attention mechanism to aggregate the historical and current neighbors of nodes, respectively. Finally, the link prediction module computes the probabilities of interactions based on the node representations after aggregating neighbors.

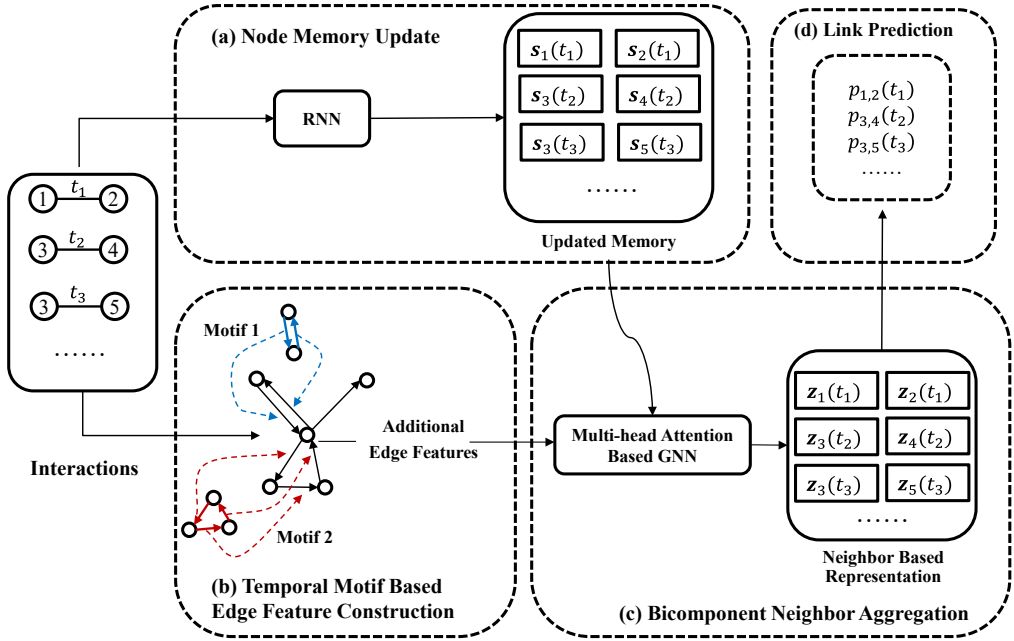

Figure 3: The framework of TME-BNA.

## 4.2 Node Memory Update

Considering that the status of involved nodes will change accordingly as an interaction occurs, we assign a memory for each node to represent the node's history [17], and utilize Gated Recurrent Units (GRUs) [32] to update the memories of nodes according to the interaction sequence. When an interaction $e = (v_i, v_j, t)$ occurs, TME-BNA takes the relevant message of this interaction to update the memories of nodes $v_i$ and $v_j$. Formally, the update operation of node $v_i$ is defined as:

$$\boldsymbol{m}_i(t) = \text{Concat}(\boldsymbol{s}_i(t^-), \boldsymbol{s}_j(t^-), \Delta_{v_i}) \tag{1}$$

$$\boldsymbol{z}(t) = \sigma_g(\boldsymbol{W}_z \boldsymbol{m}_i(t) + \boldsymbol{U}_z \boldsymbol{s}_i(t^-) + \boldsymbol{b}_z) \tag{2}$$

$$r(t) = \sigma_g(W_r m_i(t) + U_r s_i(t^-) + b_r) \quad (3)$$

$$c(t) = \sigma_h(W_c m_i(t) + U_c(r(t) \otimes s_i(t^-)) + b_c) \quad (4)$$

$$s_i(t) = (1 - z(t)) \otimes s_i(t^-) + z(t) \otimes c(t) \quad (5)$$

where $m_i(t)$ is the message of the current interaction, $s_i(t^-)$ and $s_j(t^-)$ are the memories of nodes $v_i$ and $v_j$ just before time $t$, and $\Delta_{v_i}$ is the time difference since the last interaction of node $v_i$. In addition, $W_z$, $U_z$, $W_r$, $U_r$, $W_c$, $U_c$ are weight matrices and $b_z$, $b_r$, $b_c$ are biases. In fact, $m_i(t)$, $s_i(t^-)$, $z(t)$, and $r(t)$ are the input, hidden state, update gate, and reset gate of GRU, respectively.

### 4.3 Temporal Motif Based Edge Feature Construction

The node memory update module only models the involved nodes when an interaction occurs, which cannot capture the topology of the temporal network. Inspired by GSN [31], we propose to construct additional edge features to utilize the topology of a temporal network to capture its dynamics. We extend GSN to temporal networks to capture the topology and time information jointly by counting all the instances of temporal motifs for each edge. Temporal motifs are recurrent subgraphs that can reflect the functional properties of a temporal network. Since high-order temporal motifs are hard to define and the costs of counting them are expensive, we only utilize low-order temporal motifs to construct additional edge features. According to the type of the temporal network, different temporal motifs can be used, as shown in Figure 4.

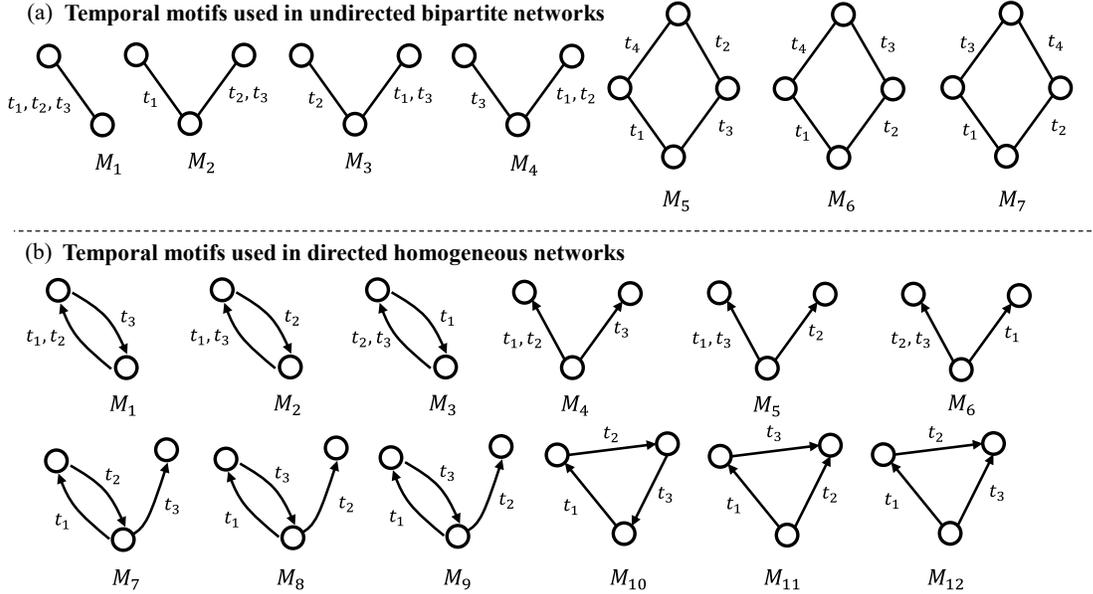

Figure 4: Different temporal motifs. (a): Temporal motifs used in undirected bipartite networks. $M_1$ represents repeated interactions, $M_2$ to $M_4$ represent common interactions, and $M_5$ to $M_7$ represent enhanced common interactions. (b): Temporal motifs used in directed homogenous networks. $M_1$ to $M_3$ represent mutual interactions between two nodes and $M_4$ to $M_{12}$ represent mutual interactions between three nodes.

If an instance of a temporal motif $M$ contains an interaction $e$, we call that the interaction forms an instance of the temporal motif. We define a set $S_M(e) = \{S_{i,M}\}_{i=1}^{N_{\text{instance}}}$ as the instances of the temporal motif $M$ that contain interaction $e$, where $N_{\text{instance}}$ is the number of instances, and $S_{i,M}$ is a certain instance of the temporal motif. We obtain the additional edge feature constructed by the temporal motif $M$ as follows:

$$r(e)_M^p = \sum_{i=1}^{N_{\text{instance}}} \text{I}(S_{i,M}^p = e), p = 1, 2, \ldots, L \quad (6)$$

where $r(e)_M^p$ is the $p$-th dimension of the additional edge feature constructed by the temporal motif $M$, and $\text{I}(\cdot)$ is an indicator function that equals to 1 if the $p$-th interaction of the instance $S_{i,M}$ is exactly the interaction $e$. Since the temporal motif $M$ contains $L$ edges, the additional edge feature based on $M$ is an $L$-dimension vector.

We construct features for all temporal motifs in the same way, and concatenate the obtained vectors to get the final edge feature of interaction $e$ as follows:

$$r(e) = \text{Concat}(r(e)_{M_1}, r(e)_{M_2}, \ldots, r(e)_{M_K}) \quad (7)$$

where $K$ is the number of temporal motifs.

Figure 5 shows an example of constructing additional edge features. The interaction $e = (v_1, v_2, 2s)$ forms the only instance $S = \{(v_1, v_2, 2s), (v_2, v_3, 5s), (v_3, v_1, 7s)\}$ of temporal motif $M_1$, and it is the interaction with the smallest timestamp. Thus its edge feature of $M_1$ is $[1, 0, 0]$.

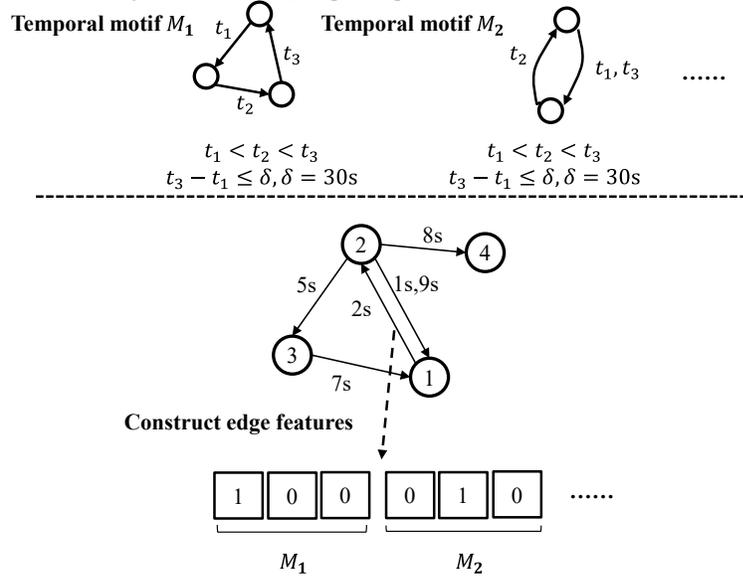

Figure 5: An example of additional edge feature construction.

### 4.4 Bicomponent Neighbor Aggregation

After obtaining additional edge features based on temporal motifs, we utilize GNNs based on the multi-head attention mechanism to jointly model the topology and time information of a temporal network. We aggregate the historical and current neighbors respectively according to the timestamps of connected edges, which can capture the topology dynamics of neighbors.

*4.4.1 Constructing the subgraphs of target nodes*

There may be repeated interactions between a certain pair of nodes with different timestamps in a temporal network, which makes the definition of neighbors different from that of a static network. In this paper, a neighbor of node $v_i$ can be represented in the form of $(v_j, t)$, where $v_j$ and $t$ represent the interacting node and interaction time, respectively. TME-BNA processes the interactions in batches, so it only needs to aggregate neighbors for the nodes involved in a certain batch. We construct subgraphs for the target nodes that need to aggregate neighbors, and then utilize GNNs on the subgraphs to obtain the neighbor-based representations of the target nodes. It is worth noting that a temporal network evolves over time, so the subgraphs of the same node at different time will change accordingly.

When obtaining the subgraph of target node $v_i$ at time $t$, TME-BNA expands the subgraph hop by hop in different ways according to the time dependence between high-order and low-order neighbors. High-order neighbors and low-order neighbors are relative, i.e., two-hop neighbors are high-order neighbors relative to one-hop neighbors while being low-order neighbors relative to three-hop neighbors. When expanding out from a low-order neighbor, denoted as $(v_{\text{low}}, t_{\text{low}})$, its neighbors can be divided into two categories:

$$N(v_{low})_{his} = \{(v_{high}, t_{high}) | (v_{low}, v_{high}, t_{high}) \in E, t_{high} < t_{low}\} \tag{8}$$

$$N(v_{low})_{cur} = \{(v_{high}, t_{high}) | (v_{low}, v_{high}, t_{high}) \in E, t_{low} < t_{high} < t\} \tag{9}$$

where $N(v_{\text{low}})_{\text{his}}$ are the neighbors of $(v_{\text{low}}, t_{\text{low}})$ reflecting its historical topology before $t_{\text{low}}$ and $N(v_{\text{low}})_{\text{cur}}$ are the neighbors of $(v_{\text{low}}, t_{\text{low}})$ reflecting its current topology between $t_{\text{low}}$ and $t$. We take the neighbors in $N(v_{\text{low}})_{\text{his}}$ and $N(v_{\text{low}})_{\text{cur}}$ as low-order neighbors, then repeat the above process iteratively to obtain the subgraph containing multi-hop neighbors.

Figure 6 shows the subgraph of node $v_1$ at time $t_8$ that contains two hop neighbors. Taking low-order neighbor $(v_4, t_5)$ as an example, its high-order neighbors can be divided into $N(v_4)_{\text{his}}$ and $N(v_4)_{\text{cur}}$ according to the time dependence with $(v_4, t_5)$. The interaction time of high-order neighbors in $N(v_4)_{\text{his}}$ is less than $t_5$, which can reflect the historical topology of $(v_4, t_5)$. While the interaction time of high-order neighbors in $N(v_4)_{\text{cur}}$ is between $t_5$ and $t_8$, which can reflect the current topology of $(v_4, t_5)$.

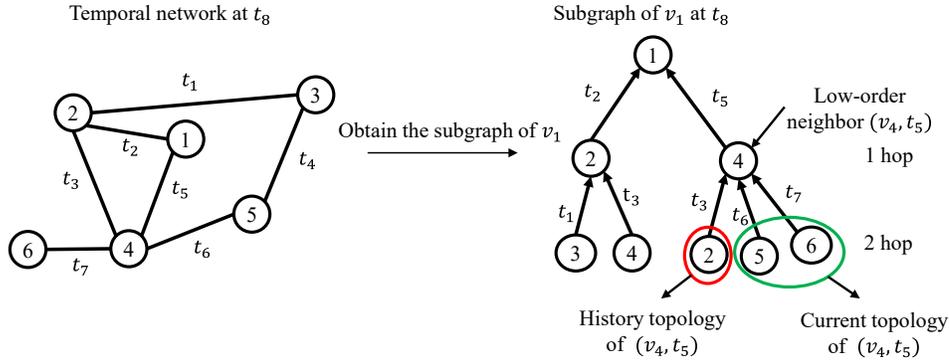

Figure 6: An example of subgraph construction.

Due to the repeated interactions, the degrees of nodes in a temporal network are usually very large, and using neighbor sampling can improve efficiency without much loss of accuracy [33]. Inspired by the performance improvement of sampling the most recent neighbors rather than sampling uniformly [17], we sample $N_{\text{neighbor}}$ neighbors in $N(v_i)_{\text{his}}$ and $N(v_i)_{\text{cur}}$ in chronological order to construct the subgraphs.

### 4.4.2 Aggregating neighbors on subgraphs

After obtaining the subgraph of a target node at time $t$, we use the multi-head attention mechanism with different parameters to aggregate $N(v_i)_{\text{his}}$ and $N(v_i)_{\text{cur}}$, respectively, which can capture the dynamics of neighbors in a temporal network. For ease of description, the notion of a neighbor node $(v_j, t)$ is replaced with $(v_j, \tau)$.

When aggregating the current neighbors $N(v_i)_{\text{cur}}$ of node $v_i$, we first use the method proposed by Xu et al. [16] to map the interaction time $\tau$ to a high-dimension vector. The mapping function is defined as follows:

$$\boldsymbol{\phi}(\tau) = \sqrt{\frac{1}{d_{\text{time}}}}[cos(\omega_1\tau), sin(\omega_1\tau), \dots, cos(\omega_{d_{\text{time}}}\tau), sin(\omega_{d_{\text{time}}}\tau)] \tag{10}$$

where $\{\omega_i\}_{i=1,2,\dots,d_{\text{time}}}$ are learnable parameters. Compared with the absolute time of the interaction, the relative time $t - \tau$ is more meaningful, so we use $\boldsymbol{\phi}(t - \tau)$ instead of $\boldsymbol{\phi}(\tau)$.

TME-BNA uses multi-head attention mechanism based GNNs to aggregate neighbors. Additional edge feature based on temporal motifs and the high-dimension vector of the interaction time are combined to obtain the query vector $\boldsymbol{q}^l(t)$:

$$\boldsymbol{q}^l(t) = (\boldsymbol{h}_i^{l-1}(t)||\boldsymbol{\phi}(0))\boldsymbol{W}_Q \tag{11}$$

where $\boldsymbol{h}_i^{l-1}(t)$ is the representation of node $v_i$ after aggregating neighbors of $l - 1$ hop and $\boldsymbol{W}_Q$ is a weight matrix.

For neighbor $(v_j, \tau)$, $e = (v_i, v_j, \tau)$ is the corresponding interaction, and the neighbor representation integrating the topology and time information is defined as follows:

$$\boldsymbol{c}_j^l(t) = \boldsymbol{h}_j^{l-1}(t)||\boldsymbol{\phi}(t - \tau)||\boldsymbol{W}_e\boldsymbol{r}(e) \tag{12}$$

where $\boldsymbol{r}(e)$ is the additional edge feature based on temporal motifs and $\boldsymbol{W}_e$ is a weight matrix. If the interactions in a temporal network are associated with feature vector $\boldsymbol{x}(e)$, we can simply extend $\boldsymbol{c}_j^l(t)$ by concatenating $\boldsymbol{x}(e)$. We stack all the vectors of neighbors in $N(v_i)_{\text{cur}}$ to get matrix $\boldsymbol{C}^l(t)$ reflecting the current topology of neighbors:

$$\boldsymbol{C}^l(t) = [\boldsymbol{c}_1^l(t), \boldsymbol{c}_2^l(t), \dots, \boldsymbol{c}_{N_{\text{neighbor}}}^l(t)] \tag{13}$$

After obtaining $\boldsymbol{C}^l(t)$, the key matrix and value matrix are defined as follows:

$$\boldsymbol{K}^l(t) = \boldsymbol{C}^l(t)\boldsymbol{W}_K \tag{14}$$

$$\boldsymbol{V}^l(t) = \boldsymbol{C}^l(t)\boldsymbol{W}_V \tag{15}$$

where $\boldsymbol{W}_K$ and $\boldsymbol{W}_V$ are weight matrices. Then, $\widetilde{\boldsymbol{h}}_{i,\text{cur}}^l(t)$, which is the representation that reflects the current topology, is obtained as follows:

$$\text{Attention}(\boldsymbol{q}^l(t), \boldsymbol{K}^l(t), \boldsymbol{V}^l(t)) = \text{softmax}(\frac{\boldsymbol{q}^l(t)\boldsymbol{K}^l(t)^T}{\sqrt{d_{\text{key}}}})\boldsymbol{V}^l(t) \tag{16}$$

$$\text{head}_p = \text{Attention}_p(\boldsymbol{q}^l(t)\boldsymbol{W}_p^Q, \boldsymbol{K}^l(t)\boldsymbol{W}_p^K, \boldsymbol{V}^l(t)\boldsymbol{W}_p^V) \tag{17}$$

$$\widetilde{\boldsymbol{h}}_{i,\text{cur}}^l(t) = (\text{head}_1||\text{head}_2||\ldots||\text{head}_k)\boldsymbol{W}^O \tag{18}$$

where $\boldsymbol{W}_p^Q, \boldsymbol{W}_p^K, \boldsymbol{W}_p^V, \boldsymbol{W}^O$ are weight matrices and $d_{\text{key}}$ is the dimension of keys. Similarly, $\widetilde{\boldsymbol{h}}_{i,\text{his}}^l(t)$, which is the representation that reflects the historical topology, is obtained. Note that the multi-head attention mechanism uses different parameters to obtain these two representations. Finally, $\boldsymbol{h}_i^l(t)$, which is the representation that captures the changes of neighbors, is obtained as follows:

$$\boldsymbol{h}_i^l(t) = f_g(\boldsymbol{h}_i^{l-1}(t)||\widetilde{\boldsymbol{h}}_{i,\text{cur}}^l(t)||\widetilde{\boldsymbol{h}}_{i,\text{his}}^l(t)) \tag{19}$$

where $f_g$ is a fully connected neural network.

The final representations of nodes can be obtained by stacking multi-layer GNNs, and the number of stacked layers needs to be consistent with the hops of the subgraphs. The representation of a node is initialized by the updated memory of the node, i.e., $\boldsymbol{h}_i^0(t) = \boldsymbol{s}_i(t)$. For ease of description, we use $\boldsymbol{z}_i(t)$ to denote the output of the last layer of the corresponding GNNs.

### 4.5 Training

TME-BNA utilizes the link prediction task for training, i.e., predicting the probability of an interaction based on the temporal network at time $t$. The probability of interaction $p_{ij}(t)$ between nodes $v_i$ and $v_j$ at time $t$ is defined as follows:

$$p_{ij}(t) = \sigma_g(f_g(\boldsymbol{z}_i(t)||\boldsymbol{z}_j(t))) \tag{20}$$

where $\sigma_g$ is the nonlinear sigmoid activation function and $f_g$ is a fully connected neural network. Since all the interactions of a temporal network are positive samples, we sample an equal amount of negative samples following the existing works [16-18]. The cross-entropy loss is defined as follows:

$$\ell = \sum_{(v_i,v_j,t)\in\mathcal{G}} -(\log(p_{ij}(t)) + \mathbb{E}_{v_n\sim P_n(v)}\log(1 - p_{in}(t)) \tag{21}$$

where $P_n(v)$ is the negative sampling distribution.

We split all the interactions into batches chronologically during training. Given a batch, TME-BNA updates node memories according to the interactions in the previous batch [17], then uses bicomponent neighbor aggregation to obtain node representations for making predictions. The detailed training process of TME-BNA is shown in Algorithm 1.

ALGORITHM 1: The training process of TME-BNA

**Input**: Temporal network $\mathcal{G} = \{e_1, e_2, \ldots, e_N\}$
**Output**: Fine-tuned TME-BNA
1. Construct temporal motif based edge features
2. Initialize TME-BNA with random network parameters
3. Split the interactions of the temporal network into $B$ batches chronologically
4. Initialize $batch\_previous \leftarrow \{\}$
5. **For** $b = 1, 2, \ldots, B$ **do:**

6.     Initialize loss $\ell \leftarrow 0$
7.    **For** each interaction in the previous batch **do:**
8.       /*Taking interaction $e' = (v'_i, v'_j, t')$ as an example*/
9.       Update the node memories of $v'_i$ and $v'_j$
10. **End for**
11. **For** each interaction in the current batch **do:**
12.      /*Taking interaction $e = (v_i, v_j, t)$ as an example*/
13.      Generate the negative sample $e^{\text{neg}} = (v_i, v_n, t)$
14.      Construct subgraphs for nodes $v_i$, $v_j$, and $v_n$, respectively
15.      Compute neighbor-based representations $\mathbf{z}_i(t)$, $\mathbf{z}_j(t)$, and $\mathbf{z}_n(t)$, respectively
16.      Calculate interaction probabilities $p_{ij}(t)$ and $p_{in}(t)$ using Equation (20)
17.      Update the cross-entropy loss $\ell \leftarrow \ell + (-\log(p_{ij}(t)) - \log(1 - p_{in}(t)))$
18. **End for**
19. Back-propagate loss and update model parameters
20. Update $batch\_previous$ with all interactions in the current batch
21. **End for**
22. **Return** fine-tuned TME-BNA

## 5 EXPERIMENTS

### 5.1 Datasets

We evaluate TME-BNA on three public datasets, i.e., UCI [34], Email-Eu [19], and Wikipedia [4]. The first two datasets are directed homogenous temporal networks and the last is an undirected bipartite temporal network. Table 1 shows the details of the used datasets.

Table 1: The statistics of datasets.

|  | UCI | Email-Eu | Wikipedia |
| --- | --- | --- | --- |
| Nodes | 1,899 | 986 | 9,277 |
| Edges | 59,835 | 332,334 | 157,474 |
| Edge feature dim | - | - | 172 |
| Timespan | 194 days | 803 days | 30 days |
| Nodes with dynamic label | - | - | 217 |

    UCI dataset is a social network of the University of California, Irvine with ~1900 nodes and ~60000 temporal edges during about half a year, where nodes are users and edges are the behaviors of sending private messages between users.

    Email-Eu dataset is a mail network of a research institution in Europe with ~1000 nodes and ~330000 temporal edges during about two years, where nodes are users and edges are the behaviors of sending mails between users.

    Wikipedia dataset is an online encyclopedia network with ~9300 nodes and ~160000 temporal edges during one month, where nodes are users and wiki pages, and edges are the behaviors of editing wiki pages. When editing wiki pages, due to some illegal operations, users will be prohibited from posting, which can be

used as labels for the node classification task. Since the edits contain text features, they are converted into LIWC-feature vectors as edge features.

## 5.2 Experimental Settings

TME-BNA is implemented based on PyTorch and Deep Graph Library [35]. The code is released on GitHub[1]. All the experiments are conducted on a Linux PC with an Intel Core i9-9900K (8 cores, 3.60G HZ) and NVIDIA RTX 2080Ti. We follow most of the experimental settings in TGAT [16] and TGN [17], and use three tasks for evaluation: transductive link prediction, inductive link prediction, and dynamic node classification.

For transductive link prediction and inductive link prediction tasks, we use the first 70% interactions for training, the next 15% for validation, and the remaining 15% for test. We randomly select 10% nodes and remove all interactions related to these nodes from the training set, and only keep the interactions related to these masked nodes in the validation and test sets for inductive link prediction. The dimensions of memory, time encoding, and node embedding are all set to 172, the number of attention heads is set to 2, the number of GNN layers is set to 2, the number of sampling neighbors during aggregation is set to 10, and we use Adam [36] optimizer and the early stopping strategy with a patience of 5 for training. In addition, we set the max length of GRU to 5 for interacting nodes in each batch to accelerate training. The above experimental settings are consistent across all datasets. For experimental settings that are different for each dataset, the choices are shown in Table 2.

Table 2: The different experimental settings for each dataset.

|  | UCI | Email-Eu | Wikipedia |
| --- | --- | --- | --- |
| $\delta$ duration | 1 day | 1 week | 1 day |
| Batch size | 100 | 200 | 200 |
| Learning rate | 0.0003 | 0.0001 | 0.0001 |

We treat dynamic node classification as the downstream task using the time-aware node representations obtained on the link prediction task. As the labels are influenced by the most recent interactions, we set the max length of GRU to 1.

As for the evaluation metrics, we use area under the ROC curve (AUC) and average precision (AP) on the link prediction tasks, and use AUC on the dynamic node classification task, as the labels are extremely unbalanced.

## 5.3 Comparison with the State-of-the-Art Methods

We compare TME-BNA with the state-of-the-art methods to show its competitive performance. A brief overview of the baseline methods is as follows:

**JODIE**: JODIE [4] employs paired RNNs to update the representations of the interacting user and item on a bipartite temporal network when an interaction occurs. Since the UCI and Email-Eu datasets are directed homogenous networks, a pair of RNNs are used to update the start and end points of the directed edges.

**DyRep**: DyRep [15] uses RNNs to update the representations of interacting nodes and their one-hop neighbors, then predicts the next interaction based on the temporal point process.

---
[1] https://github.com/pige99/TME

**TGAT**: TGAT [16] maps interaction time to high-dimension vectors used in edge features, then utilizes GNNs to aggregate multi-hop neighbors. When aggregating neighbors, all the message passing directions follow the descending chronological order.

**TGN**: TGN [17] updates the node memories in a temporal network according to the interaction sequence, and uses GNNs to aggregate the most recent neighbors of the target nodes.

**APAN**: APAN [18] updates node representations through a multi-head attention based synchronous module and propagates influence to neighbors through an asynchronous module.

For fairness, all the above baseline methods follow the same experimental settings as TME-BNA. We first evaluate the performance on the transductive and inductive link prediction tasks.

Table 3: The results of transductive link prediction on different datasets.

| Methods | UCI | | Email-Eu | | Wikipedia | |
|---|---|---|---|---|---|---|
| | AUC | AP | AUC | AP | AUC | AP |
| JODIE | 0.883 | 0.848 | 0.847 | 0.814 | 0.949 | 0.945 |
| DyRep | 0.667 | 0.616 | 0.855 | 0.830 | 0.943 | 0.948 |
| TGAT | 0.783 | 0.756 | 0.719 | 0.691 | 0.953 | 0.956 |
| TGN | 0.908 | 0.910 | 0.959 | 0.952 | 0.984 | 0.985 |
| APAN | 0.880 | 0.884 | 0.842 | 0.821 | 0.974 | 0.969 |
| TME-BNA | 0.946 | 0.950 | 0.983 | 0.980 | 0.985 | 0.986 |

Table 4: The results of inductive link prediction on different datasets.

| Methods | UCI | | Email-Eu | | Wikipedia | |
|---|---|---|---|---|---|---|
| | AUC | AP | AUC | AP | AUC | AP |
| JODIE | 0.761 | 0.743 | 0.823 | 0.808 | 0.928 | 0.932 |
| DyRep | 0.535 | 0.517 | 0.835 | 0.819 | 0.913 | 0.921 |
| TGAT | 0.684 | 0.677 | 0.593 | 0.599 | 0.936 | 0.941 |
| TGN | 0.849 | 0.862 | 0.930 | 0.920 | 0.977 | 0.978 |
| APAN | 0.895 | 0.891 | 0.837 | 0.816 | 0.977 | 0.973 |
| TME-BNA | 0.911 | 0.920 | 0.969 | 0.966 | 0.979 | 0.980 |

Table 3 and Table 4 show the performance of TME-BNA and the state-of-the-art methods on the transductive and inductive link prediction tasks, respectively. The following tendencies can be discerned:

1) TME-BNA is superior to all compared methods on all the three datasets. The AUC and AP of TME-BNA are 0.038 and 0.040 (on UCI dataset)/0.024 and 0.028 (on Email-Eu dataset)/0.001 and 0.001 (on Wikipedia dataset) higher than the best values of other methods on the transductive link prediction task, and are 0.016 and 0.029 (on UCI dataset)/0.039 and 0.046 (on Email-Eu dataset)/0.002 and 0.002 (on Wikipedia dataset) higher than the best values of other methods on the inductive link prediction task. The results demonstrate the effectiveness of TME-BNA, i.e., modeling the complex topology and time information in a temporal network jointly, and leveraging neighbor information in different time periods.

2) The baseline methods also achieve competitive performance on Wikipedia dataset, while the performance on the other datasets with large time span decreases significantly. The performance improvement of TME-BNA on the datasets with large timespan is more obvious, indicating that TME-BNA has the ability to model the long-term evolution of a temporal network.

3) The performance of DyRep on different datasets is quite different. DyRep predicts the future interactions between nodes based on the temporal point process, which is sensitive to the choice of conditional intensity functions.

4) On UCI and Email-Eu datasets, the performance of TGAT is even worse than that of JODIE, which does not utilize neighbors. The reason might be that TGAT samples the neighbors of nodes uniformly, which ignores the key fact that the most recent neighbors are more important in a temporal network.

5) On Email-Eu and Wikipedia datasets, the performance of TGN is superior to that of APAN. It might be because TGN utilizes RNNs to update node representations according to the interaction sequence, which directly captures the dynamics of nodes over time.

Table 5: The results of dynamic node classification on Wikipedia dataset.

|     | JODIE | DyRep | TGAT  | TGN   | APAN  | TME-BNA |
| --- | ----- | ----- | ----- | ----- | ----- | ------- |
| AUC | 0.875 | 0.872 | 0.858 | 0.872 | 0.835 | 0.878   |

Table 5 shows the results of dynamic node classification on Wikipedia dataset. For fairness, all baselines follow the same experimental settings as TME-BNA, i.e., treat dynamic node classification as a downstream task using the time-aware node representations obtained on the link prediction task. The results show that TME-BNA is still better than baseline methods, indicating that TME-BNA also has good generalization on downstream tasks.

## 5.4 Ablation Experiments

To verify the effectiveness of each module in TME-BNA, one of these modules is removed or modified at a time. To ensure fairness, all variants follow the same experimental settings as TME-BNA. We compare TME-BNA with the following variants:

**TME-BNA without GRU (T w/o GRU)**: This variant removes the node memory update module.

**TME-BNA without motif (T w/o motif)**: This variant removes the additional edge features based on temporal motifs.

**TME-BNA without bicompnent neighbor aggregation (T w/o bicomp)**: This variant removes the design of using the multi-head attention mechanism to aggregate historical neighbors and current neighbors respectively.

Table 6: The performance of TME-BNA and its variants on UCI dataset.

| Tasks | Metrics | TME-BNA | T w/o GRU | T w/o motif | T w/o bicomp |
| --- | --- | --- | --- | --- | --- |
| Transductive | AUC | 0.946 | 0.924 | 0.922 | 0.863 |
| link prediction | AP | 0.950 | 0.934 | 0.927 | 0.876 |
| Inductive | AUC | 0.911 | 0.897 | 0.873 | 0.837 |
| link prediction | AP | 0.922 | 0.912 | 0.886 | 0.854 |

Table 6 shows the transductive and inductive link prediction performance of TME-BNA and its variants on UCI dataset, and the following tendencies can be discovered:

1) TME-BNA outperforms T w/o GRU on all metrics. The results demonstrate that utilizing GRU to update the memories of nodes according to the interaction sequence is an effective way to model the dynamics of nodes.

2) TME-BNA outperforms T w/o motif on all metrics, indicating that introducing additional edge features based on temporal motifs is an effective way to explicitly utilize complex topology with time information. In addition, the performance improvement is more obvious on the inductive link prediction task, which demonstrates that introducing additional edge features can improve the generalization capacity of TME-BNA.

3) TME-BNA outperforms T w/o bicomp on all metrics. The results indicate that aggregating the historical and current neighbors of nodes respectively is an effective way to capture the recent changes of neighbors.

## 5.5 Parameter Sensitivity Analysis

We focus on batch size and the number of sampling neighbors in this sensitivity analysis. We evaluate the influence of above hyperparameters on the transductive link prediction task with Email-Eu dataset, and use AUC as the evaluation metric.

Table 7: Sensitivity analysis on batch size.

| Tasks | 100 | 200 | 400 | 800 |
|---|---|---|---|---|
| AUC | 0.984 | 0.983 | 0.981 | 0.978 |
| Running time of one epoch (seconds) | 413.98 | 305.82 | 252.98 | 229.25 |

Table 7 shows the performance with different batch sizes. The results indicate that as the batch size increases, the performance and the computational time of the model both decrease gradually. The reason might be that when the batch size increases, the update frequency of node memories decreases, which makes the memories out-of-date and degenerates the performance of TME-BNA. Meanwhile, increasing the batch size can reduce the computational time by increasing the parallelism of the model.

Figure 7 shows the performance with different numbers of sampling neighbors (we set the batch size to 100 due to the GPU memory limitation). The results show that the AUC of TME-BNA increases gradually as the number of sampling neighbors increases. When the number of sampling neighbors increases from 1 to 2, the performance of the model improves significantly. However, the performance of TME-BNA no longer increases when the number of sampling neighbors is larger than 10, indicating that if we sample neighbors in chronological order, sampling a small number of neighbors can obtain competitive performance. In addition, the computational time of TME-BNA increases significantly as the number of sampling neighbors increases.

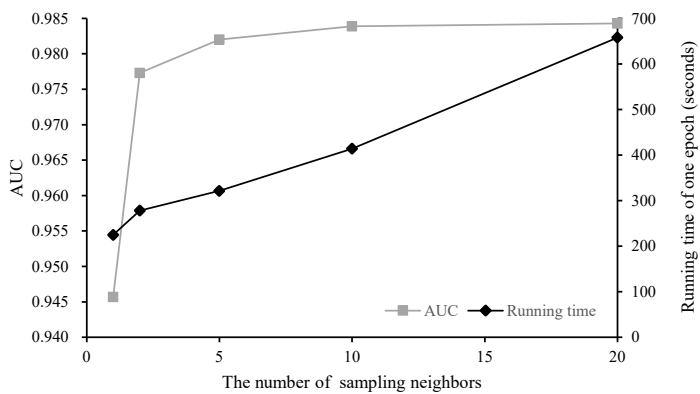

Figure 7: Sensitivity analysis on the number of sampling neighbors

## 6 CONCLUSIONS AND FUTURE WORK

In this paper, we propose a temporal motif-preserving network embedding method with bicomponent neighbor aggregation. We propose an additional edge feature construction method based on temporal motifs to explicitly integrate complex topology with time information in a temporal network. In addition, we introduce a bicomponent neighbor aggregation method based on the multi-head attention mechanism to aggregate the historical and current neighbors respectively, which can capture the recent changes of neighbors. TME-BNA is evaluated on three public datasets and the experimental results justify the effectiveness of our method.

In the future, we will extend our study in the following aspects. (1) We will improve the model to automatically distinguish important temporal motifs and assign high weights for them. (2) Since the neighbors of a node have different importance, we will introduce parameterized sampling method to construct subgraphs for neighbor aggregation.